\documentclass[%
 superscriptaddress,
 amsmath,amssymb,
 aps,
 pra,
 longbibliography,
 twocolumn,
 10pt
]{revtex4-2}
\usepackage{graphicx} 
\usepackage{physics}
\usepackage{amsmath}
\usepackage{color}
\usepackage{comment}
\usepackage[normalem]{ulem}
\usepackage{soul}

\begin{document}

\title{Efficient training of a 128-modes 3D continuously coupled photonic chip  via limited quantum resources}

\author{Denis Stanev}
\thanks{These two authors contributed equally}
\affiliation{Dipartimento di Fisica, Sapienza Universit\`{a} di Roma, Piazzale Aldo Moro 5, I-00185 Roma, Italy}

\author{Simone Di Micco}
\thanks{These two authors contributed equally}
\affiliation{Dipartimento di Fisica, Sapienza Universit\`{a} di Roma, Piazzale Aldo Moro 5, I-00185 Roma, Italy}

\author{Francesco Hoch}
\affiliation{Dipartimento di Fisica, Sapienza Universit\`{a} di Roma, Piazzale Aldo Moro 5, I-00185 Roma, Italy}

\author{Taira Giordani}
\affiliation{Dipartimento di Fisica, Sapienza Universit\`{a} di Roma, Piazzale Aldo Moro 5, I-00185 Roma, Italy}

\author{Nicol\`o Spagnolo}
\affiliation{Dipartimento di Fisica, Sapienza Universit\`{a} di Roma, Piazzale Aldo Moro 5, I-00185 Roma, Italy}

\author{Marco Gardina}
\affiliation{Ephos, Viale Decumano 34, 20157 Milano, Italy}

\author{Andrea Crespi}
\affiliation{Dipartimento di Fisica - Politecnico di Milano, Piazza Leonardo da Vinci, 32, 20133 Milano, Italy}
\affiliation{Istituto di Fotonica e Nanotecnologie, Consiglio Nazionale delle Ricerche (IFN-CNR), 
Piazza Leonardo da Vinci, 32, 20133 Milano, Italy}

\author{Francesco Ceccarelli}
\affiliation{Istituto di Fotonica e Nanotecnologie, Consiglio Nazionale delle Ricerche (IFN-CNR), 
Piazza Leonardo da Vinci, 32, 20133 Milano, Italy}

\author{Roberto Osellame}
\affiliation{Ephos, Viale Decumano 34, 20157 Milano, Italy}
\affiliation{Istituto di Fotonica e Nanotecnologie, Consiglio Nazionale delle Ricerche (IFN-CNR), 
Piazza Leonardo da Vinci, 32, 20133 Milano, Italy}

\author{Fabio Sciarrino}
\email{fabio.sciarrino@uniroma1.it}
\affiliation{Dipartimento di Fisica, Sapienza Universit\`{a} di Roma, Piazzale Aldo Moro 5, I-00185 Roma, Italy}

\begin{abstract}

\textbf{Integrated photonics provides a promising route towards the development of large-scale, stable and low-loss quantum hardware. Existing implementations of universal quantum photonic chips generally rely on architectures consisting of localised interactions between the modes that can be programmed with known algorithms to implement specific operations.
Recent advances have shown that interferometers based on 3D continuously coupled waveguide arrays  may allow to significantly scale up the number of modes with reduced overhead.
However, for this class of devices, optimal strategies to set the controllable parameters and obtain the desired transformation have not been demonstrated generally.
In this work, we  report the use of a machine learning algorithm driven by quantum data to program the operation of a reconfigurable three-dimensional continuously coupled photonic device comprising 128 modes. We show experimentally the capability of reaching a target unitary matrix, among those achievable by the device, through a limited number of one- and two-photon measurements. Our results demonstrate that this 
approach results to be effective in the presence of noise and with a limited amount of resources, also allowing to partially compensate imperfections in the experimental setup. This method can thus represent a viable route towards programming large scale integrated photonic devices,  unlocking the possibility of controlling photonic systems for which an analytical decomposition is not trivial.}

\end{abstract}

\maketitle

\section{Introduction}

The pursuit of quantum information processing has driven remarkable technological advancements, transitioning from foundational theoretical frameworks \cite{feynman_simulating_1982,DiVincenzo2000,Nielsen_Chuang_2010} to the robust experimental realization of intermediate-scale quantum devices \cite{Ransford2026, AghaeeRad2025}. A pivotal milestone 
has been the demonstration of quantum computational advantage, a regime where specific computational tasks are executed beyond the practical reach of classical supercomputers, successfully achieved using different platforms
\cite{Arute2019QuantumSupremacy,Zhong2020QuantumAdvantage,Zhong2021PhaseProgrammableGBS,Wu2021StrongAdvantage,Madsen2022QuantumAdvantage,Deng2023GaussianBosonSampling, Liu2026}. Among the various physical architectures, photonics represents a prominent platform for quantum computation, enabling versatile encoding schemes and manipulation of multi-photon states \cite{Flamini2018Review, Hasnaoui_Rev_Computing}.
In this context integrated photonics is a leading technology for quantum optical information processing, offering stable, compact and scalable circuits for the generation and manipulation of quantum states of light \cite{Wang2020IntegratedPhotonics,Pelucchi2022Roadmap,Giordani2023IntegratedPhotonics,Flamini2018Review}. Progress in single-photon sources, optical processors and detection systems has enabled increasingly large demonstrations of multi-photon interference, quantum walks and programmable linear optics \cite{Peruzzo2010QuantumWalks,Carolan2015UniversalLinearOptics,Harris2017QuantumTransport,Paesani2019GenerationSampling,Loredo2017BosonSamplingQD,Somaschi2016NearOptimalSources,Wang2019OptimalSources}. In particular, Boson Sampling represents a central benchmark for photonic quantum technologies, linking multi-photon interference in random interferometers to computationally hard sampling problems \cite{Aaronson2011ComputationalComplexity,Brod2019BosonSamplingReview}. This paradigm has driven experiments from early few-photon implementations to large-scale Gaussian Boson Sampling demonstrations of quantum computational advantage \cite{Broome2013PhotonicBosonSampling,Spring2013BosonSamplingChip,Crespi2013IntegratedBosonSampling,Tillmann2013ExperimentalBosonSampling,Zhong2020QuantumAdvantage,Madsen2022QuantumAdvantage,Deng2023GaussianBosonSampling}. At the same time, programmable photonic processors are also attracting interest for quantum simulation, quantum chemistry, graph problems, randomness generation and machine-learning-inspired photonic computing \cite{Harris2017QuantumTransport,Huh2015Vibronic,Arrazola2018DenseSubgraphs,Schuld2020GraphSimilarity,Banchi2020MolecularDocking,Shi2024BosonSamplingRandomness,Hoch2024BernoulliFactory,Steinbrecher2019QuantumOpticalNN,Yan2024ReservoirComputing, ClementsTrainabilityPhotonicQML}.

Most reconfigurable linear-optical processors rely on meshes of discrete beam splitters and phase shifters, where universal decomposition algorithms provide a direct mapping between a target unitary and the circuit settings \cite{Reck1994Universal,Clements2016Optimal}. This guarantees programmability, but scaling to many modes increases circuit size, loss and calibration complexity \cite{Carolan2015UniversalLinearOptics,Taballione2019ReconfigurableProcessor,Arrazola2021ProgrammableNanophotonic}. Continuously-coupled waveguide arrays offer an alternative architecture, in which photons evolve through a lattice of mutually coupled waveguides over the full propagation length, and the implemented transformation can be described by an effective Hamiltonian \cite{Peruzzo2010QuantumWalks,Hoch2022Reconfigurable32,DenisCCSimTrain}. The use of three-dimensional femtosecond-laser-written circuits enables compact layouts with enhanced connectivity and reduced circuit length, making this approach attractive for large-scale low-loss photonic processors \cite{Gattass2008Femtosecond,Arriola2013LowBendLoss,Hoch2022Reconfigurable32}. This strategy has been demonstrated in a 32-mode reconfigurable continuously-coupled 3D circuit for Boson Sampling \cite{Hoch2022Reconfigurable32}, and recently extended to a 128-mode reconfigurable processor \cite{DiMicco2026BosonSampling128}. In this device, longitudinal modulations of the waveguide positions generate static randomness in the evolutions, while thermo-optic micro-heaters provide active reconfiguration by modifying the propagation constants. 
The main open challenge for this class of processors is programmability \cite{Yang_CCW_Rev}. 
Previous works \cite{youssry_2025} have shown an approach to achieve universal programmability in continuously coupled interferometers, which requires circuits of sufficient depth and reconfiguration capabilities. However, this technique is beyond the reach of currently available devices due to the high requirement in terms of reconfigurability and size.
Unlike universal mesh interferometers, continuously coupled devices lack a general analytical recipe to translate a desired unitary into experimentally accessible heater settings. Additionally, devising a calibration method based on full tomography over many configurations becomes impractical at large scale. However, many near-term applications only require reproducible access to useful transformations or the optimization of task-specific output statistics. This motivates black-box, measurement-based training strategies, where the chip is treated as a trainable physical system \cite{Youssry2020DeepLearning,DenisCCSimTrain}. In this context, quantum data-driven optimization techniques are particularly well-suited to infer effective control settings directly from experimental measurements, bypassing the need for a fully characterized analytical model of the quantum processor \cite{Zhang2020QuantumMLReview, Bartlett2020ProgrammablePhotonicArchitecture, Youssry2020DeepLearning,DenisCCSimTrain}. In particular, the combination of machine learning and quantum technologies has recently emerged as a promising research direction, finding applications in quantum information processing, photonic computing, circuit optimization and quantum-enhanced learning tasks \cite{Stanev2023QuantumOpticalCloning,Steinbrecher2019QuantumOpticalNN,Zhang2020QuantumMLReview,Bordoni2023LLPAnomalyDetection,Bartlett2020ProgrammablePhotonicArchitecture,Suprano2024PhotonicELM,Hoch2025AdaptiveBosonSampling,Nielsen2024ProgrammableNonlinearPhotonics}.

\begin{figure}[h!]
    \centering
    \includegraphics[width=1\linewidth]{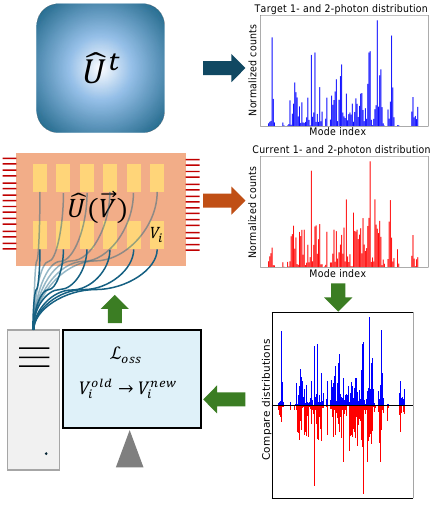}
    \caption{\textbf{Conceptual figure of the training process.} Our objective is to train the optical chip to perform a specific desired unitary transformation $U^t$, using only a limited number of single- and two-photon measurements. To this end, we prepare states of single-photon and of two indistinguishable photons at the input of the chip and collect the output photon distributions. The tunable parameters are the voltages $V_i$ applied to the resistances above the photonic circuit, which in turn alter the propagation constants of the waveguides due to thermo-optic effects. For this process, we first obtain the target one- and two-photon photon distributions from the target unitary. Then, we use machine learning techniques to adjust these parameters over multiple epochs to obtain a result that is as close as possible to our target. For each epoch, we sample the output distributions from the chip with the chosen parameter values, then compare them to the target distributions defined at the beginning, and calculate the loss, which we then use to update the values of the parameters $V_i^{old}\rightarrow V_i^{new}$. Those new parameter values are then fed back to the optical chip, and the cycle repeats, until a chosen number of epochs has passed.}
    \label{fig:conceptual}
\end{figure}

Here, we experimentally investigate a measurement-based, quantum data-driven machine learning approach \cite{DenisCCSimTrain} to program a femtosecond laser-written continuously coupled waveguide array comprising 128 modes. The device is actively controlled by up to 18 resistors, and we show how the employed method is capable of finding the optimal electrical currents required to program a target unitary transformation among those achievable by our device. This goal is achieved using a limited number of one- and two-photon measurements, without the need for extensive characterisation or a precise modelling of the circuit's internal functioning. An overall representation of the experimental setup and of the training process is shown in Fig. \ref{fig:conceptual}.
We employ multiple optimisations to reduce the training time and increase its stability against typical noise sources of single-photon integrated hardware, such as the finite statistics of single-photon counts, photon partial distinguishability, and the thermal instability of the heaters.
We successfully demonstrate the capability of programming the chip operation within a limited timeframe.

The remainder of this paper is organised as follows. First, we provide a brief theoretical overview of the continuous-coupling paradigm. Then, we describe the integrated photonic chip employed in the experiment, followed by an overview of the experimental platform and of the training algorithm used to reconfigure the device.
After analysing the physical limitations and noise sources of the setup, we demonstrate the capability of the algorithm to converge towards a random reachable target unitary from independent starting heater settings.
We then demonstrate a more specialised use case of the algorithm, showing how it can be employed to recover a target chip configuration after performance drifts caused by imperfections in the single-photon source or by issues with the controllable {heaters}. Finally, we discuss the overall results, the current limitations of the approach, and possible future developments.

\begin{figure*}[t]
    \centering
    \includegraphics[width=1\linewidth]{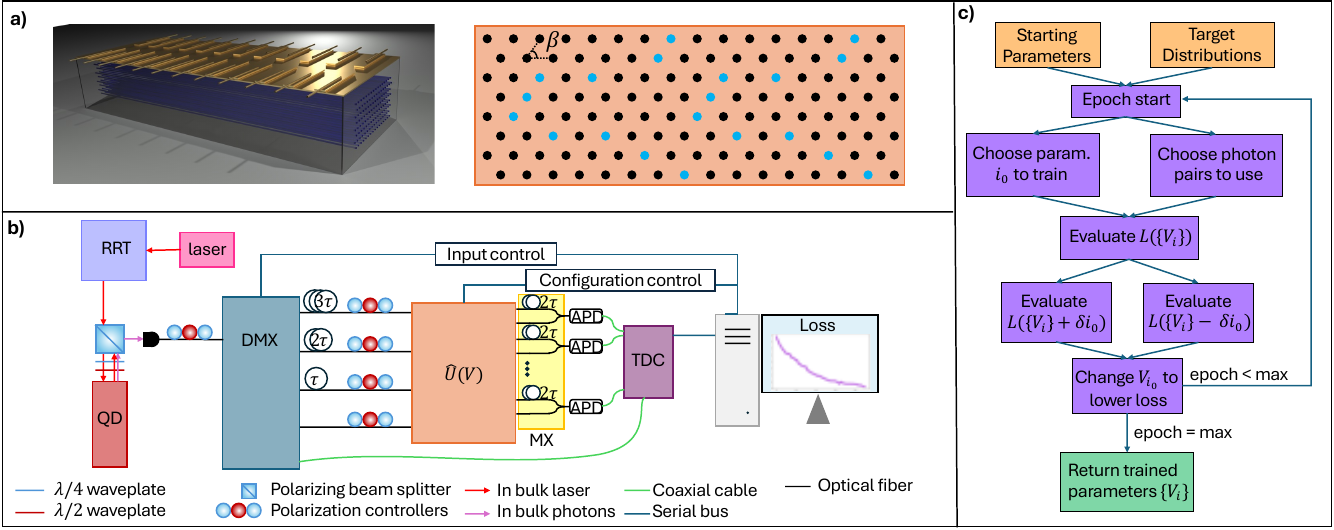}
    \caption{\textbf{Conceptual and experimental scheme of the experiment.} In panel \textbf{a}, we show a 3D rendering of the reconfigurable photonic integrated chip. On the top surface, the micro-heaters used to generate the temperature gradient through thermal dissipation upon voltage application are represented. A cross-sectional view of the interference region is also shown, illustrating the arrangement of the 128 waveguides. The waveguides are arranged in a triangular lattice consisting of 16 columns and 8 rows. The vertical spacing is smaller than the horizontal spacing, resulting in a lattice composed of isosceles triangles with a base angle of $\beta=56^\circ$. The available input modes are shown in blue. In panel \textbf{b}, we show a schematic of the Qolossus 3D platform. The pump laser is first sent to a repetition-rate tripling (RRT) system, consisting of a Mach-Zehnder interferometer followed by a Michelson interferometer. The laser then excites the quantum dot (QD), passing through a cross-polarisation setup that separates the emitted photons from the reflected pump light. The photons pass through polarization controllers to maximize the efficiency of coupling into the time-to-space demultiplexer (DMX). At the output of each DMX arm, the photons are synchronised using delay lines and subsequently pass through additional polarisation controllers to ensure polarization indistinguishability at the chip input. At the chip output, pairs of output modes are combined into a space-to-time multiplexer (MX). Specifically, for each pair of output modes, one mode is delayed before both are injected into a multimode fiber with a core large enough to simultaneously collect light from both outputs. This scheme reduces by half the number of single-photon avalanche photodiodes (APDs) required for detection. The APDs are connected to a time-to-digital converter (TDC), which records the time tags of all detection events. The TDC also acquires a trigger signal from the DMX, which is required to identify the multiplexed modes based on their arrival times relative to the trigger. Finally, the data is processed in real time by a classical computer, which registers the single- and two-photon counts for each mode and executes the training algorithm by computing the loss function, selecting the input modes, and reconfiguring the chip. In panel \textbf{c}, we show a schematic representation of the implemented training algorithm.}
    \label{fig:exp_fig}
\end{figure*}
 
\section{Methodology}

\subsection{Chip Structure}

In this work, we employed a reconfigurable integrated continuous-coupled interferometer comprising 128 optical modes arranged in a triangular lattice of 16 columns and 8 rows \cite{DiMicco2026BosonSampling128}, as shown in Fig.~\ref{fig:exp_fig}a. The chip is fabricated in borosilicate glass, and the waveguides are realised using the femtosecond laser writing technique. The distance between adjacent waveguides is subject to a slight random modulation in order to prevent the interferometer from exhibiting predictable behaviour arising from symmetry considerations.

At the input facet of the chip, 20 waveguides are accessible through permanently attached single-mode fibers. At the output facet, a fan-out region rearranges the waveguides into a square grid. A detachable fiber array then couples all 128 modes to an equal number of multimode fibers in order to minimise coupling losses. On the top surface of the interferometric region of the chip, 24 resistive micro-heaters are fabricated, with 12 located on the left side and 12 on the right side, as shown in Fig.~\ref{fig:exp_fig}a.
By applying a controllable tension up to $8$V, the heaters induce local temperature gradients that modify the refractive index of the substrate through the thermo-optic effect, thereby changing the propagation constants of the waveguides and making the device reprogrammable.

However, since a complete and detailed characterisation of the chip would be unfeasible due to its size, we do not have access to an analytical formula to select the correct voltages to apply to the micro-heaters to perform a desired unitary transformation.

\subsection{Experimental Setup}
A scheme of the complete setup is shown in Fig.~\ref{fig:exp_fig}b. We employ a Quantum Dot (QD) single-photon source to generate a stream of indistinguishable single-photons. The QD is resonantly pumped using a pulsed laser at 927 nm with a repetition rate of $80$ MHz. To triple the laser repetition rate, we utilise a combination of an unbalanced Mach-Zehnder and a Michelson interferometer. This results in an effective pump repetition rate of 240 MHz. Since the source is resonantly pumped, we employ a cross-polarisation system to separate the reflected laser from the single-photons. This allows us to select the photons emitted in a polarisation orthogonal to the pump laser. The apparatus has an extraction efficiency of $\sim10\%$. The single-photon purity {$g^{(2)}(0)$} of the source is $\sim4\%$, and the single-photon indistinguishability {of subsequently emitted photons is $M_s \sim 97 \%$. We calculated it as $M_s=(V_{HOM}+g^{(2)}(0))/(1-g^{(2)}(0))$ \cite{Ollivier2021}, where $V_{HOM}$ is the Hong–Ou–Mandel visibility.} 
The stream of photons is directed to a time-to-space demultiplexer system (DMX). The DMX redirected the photons to up to four different paths, generating trains of photons with a length of approximately 200 ns. The four outputs of the DMX are delayed to synchronise the streams. Additionally, they pass through paddle polarisation controllers (PPC) to compensate for the polarisation rotations of the four different paths. Finally, each output of the DMX is connected to an input mode of the chip. To check the polarisation of the input photons, we detach the fiber array at the output of the chip and place a polarising beam splitter (PBS); the reflected component is then minimised using the PPCs. 
When injecting the interferometer with quantum input states, the output multimode fibers are connected to a custom space-to-time multiplexing system, which combines two spatial modes into one, after delaying one of the two. This process maps spatial modes to time bin modes, effectively halving the number of detectors required. The detection is carried out using Single-Photon Avalanche Diodes (SPAD), with a detection efficiency of $\sim 30\%$ at the wavelength of 927 nm and a jitter time of $\sim 350$ ps FWHM.
The SPADs are connected to time-to-digital converters (TDCs) to record the single-photon events and their arrival times. We used two TDCs, each with 32 channels, a resolution of 1 ps, and a jitter time of 20 ps. The internal clocks of the TDCs are synchronised using a dedicated wired connection, and their initial times are synchronised by splitting the signal from a single detector and connecting it to both TDCs. Additionally, a trigger signal provided by the DMX is connected to one channel of one of the two TDCs. This signal serves as a reference to distinguish the time bin modes of the multiplexing system. Finally, custom software is employed to post-process the signal and extract multi-photon coincidence events.
In this work, we use a total of 51 SPADs, resulting in the measurement of 102 modes out of the available 128. Initially, we employed 18 micro-heaters; subsequently, one was showing instabilities and was disconnected, thus we used the remaining 17.

\subsection{Training}

We now describe the training procedure for our method, which is based on the training procedure described in the work of Ref.\cite{DenisCCSimTrain}, where it was tested via numerical simulations. A schematic representation of the training procedure can be seen in Fig.~\ref{fig:exp_fig}c .

Our objective is to train the optical chip to perform a specific desired unitary transformation $U^t$, using only a limited number of single- and two-photon measurements. {To this end}, we treat the optical chip as a machine learning model, where the inputs are the single- and two-photon states we send into the chip, the outputs are the output photon distributions, and the tunable parameters are the voltages $V_i$ applied to the resistances, which in turn alter multiple propagation constants due to thermo-optic effects. We then use machine learning techniques to adjust these parameters over multiple epochs to obtain a result that is as close as possible to our target.

We chose to use the squares of the voltages $V_i^2$ as tunable parameters, instead of using the voltages $V_i$, because that gives a behaviour that is closer to being linear with respect to the propagation constants and the loss. More specifically, the shift on the propagation constants depends on the shift in temperature, which in turn depends on the shift of the dissipated power, which itself depends on the square of the voltage applied, assuming a constant resistance. While this is a simplified model, we do find that the behaviour of the loss function is close to linear with respect to $V_i^2$, as we can see from the loss landscapes shown in Appendix \ref{app:LossLandscape}.

A technique that is often used in machine learning is backpropagation \cite{Backpropagation}, which allows to quickly obtain the gradient of the loss, and thus determine how to update the values of the parameters. However, backpropagation requires a complete characterization of the device to compute the necessary derivates, and thus such a technique cannot be applied in our scenario. This means that we have to resort to alternative ways to approximate the gradient and decide in which direction to shift the parameters. 

In our approach, we used a modified version of Finite Differences Stochastic Approximation (FDSA) \cite{FDSA1, FDSA2}, where at each epoch we choose a single parameter. 
The way the parameter to train is chosen at each epoch is the following: all the usable parameters are organized in a randomly ordered list, and for each epoch we select the next parameter in the list. Once we reach the end of the list, meaning that the full set of parameters has been used, the list is randomly shuffled to obtain a different random ordering, and we repeat the cycle starting again from the beginning of the list. This method, while retaining a degree of randomness in the order the parameters are trained, ensures at the same time that no parameter is left untouched for too long. 
The way our modified FDSA works is that, at each epoch, after having chosen the parameter to train for that epoch, we measure the loss for the current configuration of the chip, then proceed to shift the value of that parameter up and down by a certain value $\delta_{shift}$, measuring the loss in each case. Thereupon, we move in the direction of lower loss. At this point, we can either take a step proportional to the gradient of the loss, or set the parameter directly to the one measured with lower loss. In our case we found out the second option to be more reliable, as it was less sensitive to noise when close to the target. The value we choose for $\delta_{shift}$ needs to be a compromise between precision and noise resistance. Indeed, the smaller the value, the more theoretically precise the result, but, at the same time, the more the algorithm is sensible to noise. Below a certain threshold the oscillations of the experimental setup become comparable with the change in loss caused by the shift in the parameter, making it impossible to reliably determine the right direction. In our case we chose $\delta_{shift}$ to be proportional to the chosen learning rate, which we reduce during the training as epochs pass, and to the loss of the starting position of the current epoch, thus allowing for more precision when closer to the target. For {more details on the reasoning behind some of the choices we made for our modified version of FDSA}, the reader can refer to \cite{DenisCCSimTrain}, {where this method was applied on multiple simulated architectures}.

At the beginning of the training process, the chip is set to an initial starting configuration, independent of the target. We then provide the training algorithm with the target one- and two-photon distributions, which will be used to evaluate the loss function at each step, by comparing the target distribution with the measured output distributions, as we explain in more detail later. The training loop then consists of multiple epochs; at each epoch we choose a parameter to train, and which input pairs to use for the epoch. We then proceed to use the modified FDSA method described above, using single and two-photon measurements, to choose in which direction to shift the parameter to minimise the loss. After the value of the parameter is changed, the epoch concludes, and we start the next epoch, where we repeat the cycle with a different chosen parameter and a different set of input pairs. We repeat this cycle for a predetermined number of epochs, after which the training concludes.

For the loss function $\mathcal{L}(\{ V{i} \})$ we use the Mean Absolute Error (MAE) between the measured samples $\tilde{y}$ and the target distribution $y^{t}$, calculated as:
\begin{equation} \label{eq:loss_func}
\begin{aligned}
    \mathcal{L}(\{ V{i} \}) &= \frac{1}{\#M_1}\sum_{(i,j)\in M_1} \vert \tilde{y}_{i;j} - y^{t}_{i;j} \vert +\\
    &+  \frac{1}{\# M_2}\sum_{(i,k,j,l) \in M_2} \vert \tilde{y}_{i,k;j,l} - y^{t}_{i,k;j,l} \vert,
\end{aligned}
\end{equation}

where $M_1$ is the set of the measured single-photon input-output combinations $(i,j)$, $M_2$ is the set of measured two-photon input-output combinations $(i,k,j,l)$,  and $\#M$ refers to the cardinality of the set $M$.
The target distributions have been previously obtained by sampling from the chip with the target parameters set for a longer period of time. This is to ensure that the target is reachable, but these target distributions can also be obtained in theory from a desired target unitary through simulation. However, if one wishes to simulate the target output distributions, they would also have to take into account the imperfections of the experimental setup, such as those caused by the single-photon source, in said simulations. The loss function we use gives the same weight to the MAE of the one and two photon distributions. We found empirically that this balance between the two parts of the loss works well in our case, as the one-photon loss tends to be less effected by statistical fluctuations, and thus provides a baseline stability, while the two-photon loss is needed to have sensitivity to the phases of the unitary. While in our case this balance worked well, in some other tasks one may wish to weigh the contributions of the one and two photon components differently. Note that using this loss function, which directly compares the single and two-photon distributions, is significantly less expensive in terms of measurement time than comparing the target matrix and the one performed by the chip at each step, which should be reconstructed each time.

We implemented several optimisations to further reduce training time, while minimising the impact on training quality. Since we {make use of} 4 input modes, we have 4 possible single-photon input configurations, and 6 possible two-photon input configurations. To optimise the training time, at each epoch we use only 3 of the 6 possible two-photon input configurations. This allows us to cut the training time almost in half while maintaining a similar quality of the final trained result, since the time needed to measure the two-photon inputs is significantly longer than the time needed to measure the single-photon inputs.

Due to the generation rate of photons, and the large number of modes present in the chip, to obtain a sample of measurements that is statistically robust, we need to perform measurements for a sufficient amount of time. The amount of time needed for acquiring significant two-photon statistics is significantly longer than that needed for single-photon statistics, due to the additional losses and the increased number of possible output combinations. To be more specific, in our experimental setup we measured, for single-photon inputs, approximately $5 \times 10^5~\mathrm{Counts/s}$, while for two-photon inputs we measured approximately $3 \times 10^3~\mathrm{Coincidences/s}$. In the single-photon case, the number of possible output configurations is simply equal to the number of measured output modes, namely 102. In contrast, for two-photon inputs, the number of possible output configurations is given by the binomial coefficient $\binom{102}{2}=5151$. Therefore, the measurement time must be chosen to ensure that sufficient statistics are collected for all output configurations.

In our tests, we found out that using 6 seconds of measurement time for each single-photon input, and 60 seconds for each two-photon input, was a good compromise between measurement time and quality of the result. In the case of single-photon measurement, we found that we could also use 1 second measurements per input, and obtain comparable results, but reducing the measuring time for the two-photon inputs led to measurably worse results. {In Appendix \ref{app:LossLandscape} we show a comparison of the loss for the tested measurement durations.} To speed up the training, while maintaining the quality of the final trained result, we employed a two-step training process. First, we pass through a quick ``rough" training phase, where we use only single-photon measurements for 1 second each and a high learning rate, to find the area near the minimum. Then, we proceed to do a second longer training phase, using both single- and two-photon measurements, with 6 and 60 seconds for each input respectively, with a lower learning rate{, and using learning rate decay}. This allows us to spend less time on the initial part of the training, where less precision is required, and spend more time optimising when close to the target result. In this way we can maximise the quality of the result within a limited time-frame.

{We found that} the method was still capable of successfully training the optical chip to perform the desired transformation, within a certain margin of error, {even in the presence of moderate experimental noise,} as we will see in more detail in the next section.

\section{Results}

To evaluate the quality of the trained result, we use the entry-wise $L^1$ norm of the difference between the matrix obtained with the trained parameters and the target matrix:

\begin{equation}
    L^1_{norm} = \frac{1}{2} \sum_i^n \sum_j^m |a^p_{ij} - a^t_{ij}|,
\end{equation}
where $a^p_{ij}$ represent the elements of the complex matrix obtained with the trained parameters, and $a^t_{ij}$ the elements of the target matrix, with the sum being done over the $n$ measured input modes and the $m$ measured output modes. We choose to use this metric since we cannot use Fidelity, as what we can reconstruct is not the full unitary matrix, but only a $4 \times 102$ submatrix, due to the limited number of inputs and outputs we have access to.

In Appendix \ref{app:corr_Frob_Fid} we show how the $L^1_{norm}$ and the Fidelity are correlated, to give a better point of reference for the performance of the training method.

\subsection{Noise and Stability} \label{sec:noise_and_stability}

\begin{figure*}[t]
    \centering
    \includegraphics[width=1\linewidth]{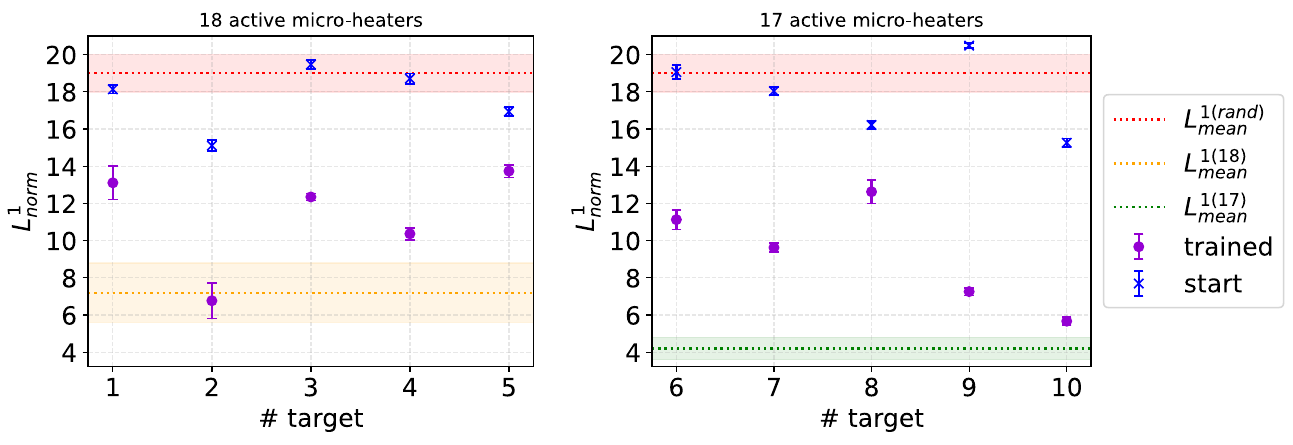}
    \caption{\textbf{$L^1_{norm}$ of the performed trainings.} In this figure, we report the $L^1_{norm}$ of the difference between the target matrices and both the matrices of the starting point (crosses) and the matrices obtained after training (dots). The red dotted line indicates the average $L^1_{norm}$ distance between random matrices implemented on the chip, while the shaded region represents one standard deviation. The yellow (green) dotted line indicates the average $L^1_{norm}$ distance among multiple reconstructions of the same chip configuration with 18 (17) active micro-heaters, and the corresponding shaded region represents one standard deviation. Hence, they show the minimum $L^1_{norm}$ reachable due to the noise of the setup, and are obtained as described in Section \ref{sec:noise_and_stability}. The uncertainty bars are computed using a Monte Carlo procedure, as explained in Appendix~\ref{app:ErrorComputation}.}
    \label{fig:norm_target}
\end{figure*}

{Before discussing the results, we first need to address the main sources of error and uncertainty that were intrinsic to the experimental setup, and not dependent on the training algorithm. This is needed to obtain a baseline for what the optimal result would look like, as well as to better understand the challenges faced by the training algorithm.}

{We focused on three main sources of uncertainty: the statistical uncertainty arising from the use of a limited number of output samples, the single-photon source fluctuations, and imperfections in the chip. The contribution to the uncertainty from these three sources cannot be cleanly disentangled; therefore, in the following analysis, we provide a comprehensive discussion aimed at assessing their combined impact on the measurements and on the reconstructed matrices.}

The algorithm used to reconstruct the unitary submatrix is the one described in \cite{laing2012}. {This} method relies on single-photon measurements to determine the moduli of the matrix elements and on two-photon visibility measurements to estimate their phases. {We highlight that this reconstruction algorithm is insensitive to a layer of phases of the matrix in input and output. However, since the $L^{1}_{norm}$ is sensitive to these phases, we made sure in the reconstruction algorithm to set these indeterminate phases to the same value, so that they would not affect the comparison of the reconstructed matrices.} Experimentally, the two-photon visibility is obtained from the ratio between the central peak and the side peaks in a time-correlation histogram. As a result, the accuracy of this measurement strongly depends on the amount of collected statistics. 

As a first test, we sequentially measured the same submatrix three times. We will refer to this configuration, with 18 active resistors, as the reference target $T_{ref}^{(18)}$. We then compared the reconstructed submatrices using the $ L^1_{norm}$ between these measurements. {Each of the 4 single photon inputs was measured for 10 minutes, while} each {of the 6} input pairs was measured for {20} minutes. {Including the required post-processing time, this led} to an overall measurement time to reconstruct a single submatrix of $\sim4$ hours. We obtained an average distance between the reconstructed submatrices of $L^{1(18)}_{mean}=7.2 \pm 1.6${, where the uncertainty comes from the dispersion of the measurements, i.e. $\sigma = \sqrt{\sum_i^n(\mu-x_i)^2/n}$, where $\mu$ is the mean value, $x_i$ are the single measurements and n is the number of measurements}. 
 
Subsequently, we measured the same target on different days using 17 active resistors, again with a measurement time of twenty minutes per input pair. Taking the first measurement, denoted as $T_{ref}^{(17)}$, as a reference, we computed the distance of all subsequent reconstructions with respect to it and obtained an average distance of $L^{1(17)}_{mean}=4.3\pm0.7$. This result suggests that measurements performed with 17 active resistors are more stable, even over longer time periods. We believe that the larger fluctuations observed in the 18-resistor configuration were caused by {the instability of the 18th resistor.} 

Moreover, we reconstructed the submatrix on four different non-consecutive days with all resistors turned off. Comparing the reconstructed matrices, we obtained an average distance of $L^{1(0)}_{mean}=3.7\pm0.5$, which is comparable to the value obtained for repeated measurements of $T_{ref}^{(17)}$. This result indicates that the heaters exhibit good long-term stability.

The latter result also provides an estimate of the overall long-term stability of the system, provided that the conditions of the single-photon source are carefully monitored and restored before each measurement. In particular, it is necessary to recover approximately the same single-photon purity and emission rate. {This is because, when we define the targets, the source imperfections play a role in the final output statistics. Hence, in order to be sure that the applied unitary transformation is the same, we also have to apply the same experimental conditions. To assess how the {overall} state  {of the experimental setup} affects the final output distribution, and hence, our ability to reconstruct the applied unitary matrix} we  {leave} the same chip configuration {for} several days after the last optimisation of the source conditions. In this case, comparing the result with $T_{ref}^{(17)}$, we obtained a distance of $L^1_{norm}=9.3$, {showing that the system needs a periodic recalibration to properly work.}

Finally, we investigated the effect of the collected statistics by doubling the measurement time and sequentially measuring the same configuration $T_{ref}^{(17)}$ three times. Comparing these reconstructions with one another, we obtained an average distance of $L^{1(long)}_{mean}=3.6 \pm 0.6$. These measurements clearly show how the reconstruction quality depends on the acquisition time, and therefore on the amount of collected statistics. Nevertheless, {when comparing this result with }the twenty-minute measurements {we find that, while the forty-minute measurements show slightly fewer fluctuations, they are still compatible. Hence, 20 minutes are} already sufficient to validate the training procedure, and we therefore retained this acquisition time for efficiency reasons.\\
In Appendix \ref{app:LossStability} we report how the short-term stability of the system affects the loss function defined in Eq.~\eqref{eq:loss_func}, while in Appendix \ref{app:LossLandscape} we report the loss landscape of a single parameter when using different measurement times. These additional analyses allow us to better understand how these sources of noise due to both the source and the limited statistics, affect the loss and, as a result, the training.

\subsection{Reaching random unitaries} \label{sec:Random_Unitaries}

\begin{figure*}[t]
    \centering
    \includegraphics[width=1\linewidth]{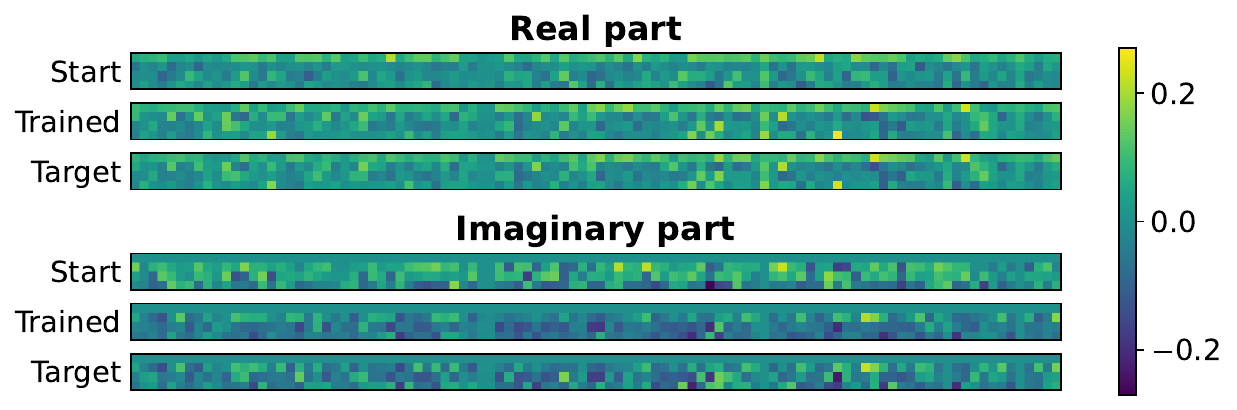}
    \caption{\textbf{Comparison of the reconstructed matrices} In this figure, we compare the reconstructed matrices of a training for the starting point of the training, the trained matrix, and the target matrix. The training used was that of the target 9 from Figure \ref{fig:norm_target}.}
    \label{fig:MatrixCompare}
\end{figure*}

First, let us briefly discuss the chosen training time. We optimised the training for a fixed time frame, and in our case we decided to limit the training time to around 1 day, which is approximately equal to the measurement time needed to reconstruct 6-7 unitaries at the sampling rate of our experimental setup which, as we explained in the previous section, takes $\sim4$ hours each. We chose this training time as we found that, in our tests, it was a good compromise between measurement time and quality of the result. We achieved this by implementing the optimisations described in the Methodology section, including having a first faster training phase using only single photons, followed by a second longer training phase using also photon pairs, using only 3 of the 6 possible photon pairs for each epoch, and optimising the measurement time per input to achieve a balance between speed and noisiness. 

In order to test the effectiveness of the algorithm we tested its performance by attempting to reach 10 different random target configurations of the chip. The target configurations have been randomly chosen by selecting random values for the tensions applied to the micro-heaters, and then characterising the corresponding sub-matrix describing the unitary transformation performed by the chip. This is done to ensure that the target configurations are reachable by the chip. For all trainings, the starting position was the same, as to be target-independent, and was chosen to be the one with all values of ${V_i^2}$ in the middle of the usable range. As we previously mentioned, {the first} half of the training was done using 18 heaters, while the other half was done using 17 heaters.

The training was performed as described in the Methodology section, using the two-step training process. For the first training step, the one with only single-photon inputs, we trained for 90 epochs in the case of 18 heaters and 85 epochs for 17 heaters. For the subsequent training step, we trained using both one- and two-photon inputs, using only 3 two-photon inputs per epoch, and training for an additional 90 epochs in the case of 18 heaters, 85 epochs for 17 heaters. 
The number of epochs was chosen to be a multiple of the number of trainable parameters, so that each parameter gets modified the same number of times, since we adjust only a single parameter at each epoch. {For the first training step we used a high learning rate, while for the second training step we used a lower initial learning rate, and made use of learning rate decay.} {In Appendix \ref{app:LossTraining} we show the behaviour of the loss during the training process.}

In Fig.~\ref{fig:norm_target} we report the values of the $L^1$ norm between the target sub-matrices and the reconstructed sub-matrices obtained by applying the tensions returned at the end of training. To give a reference point, the average $L^1_{norm}$ between unitaries sampled randomly from the chip is $L^{1(rand)}_{mean} = 19.02 \pm 1.22$ . In appendix A we discuss the relation between the $L^1_{norm}$ of sub-matrices and the fidelity of the complete unitary matrices.

From the results, we can see that, while the quality of the trained result fluctuates, the training algorithm is always able to improve the loss with respect to the starting configuration, in some cases even coming close to the limit of distinguishability given by the noise of the experimental setup. We do also see an increase in the trained performance after we started using only 17 heaters instead of 18. That is likely because the {18th heater} could have been less stable than the others, as we also saw in the stability measures discussed previously, and thus has led to a worse trained result. In Fig. \ref{fig:MatrixCompare} we can also see a direct comparison of the reconstructed matrices for one of the trainings, better visualising how close our trained matrix can come to the target one.

\subsection{{Training towards an unreachable target}}

Next, we examined how our training algorithm performs in a situation in which the target it needs to reach is not achievable due to limitations of the device. The way we tested this scenario was by attempting to train towards targets obtained using 18 heaters, while making use of only 17 of the heaters. The training procedure employed was the same as the one used for training towards random unitaries, including using a starting position that is target independent.

\begin{table}[ht]
\begin{center}
\begin{tabular}{|c | c | c | c | c | c |} 
 \hline
 Target num. & $T_{1}$ & $T_{2}$ & $T_{3}$  \\
 \hline
 $L^1_{norm}$ Trained &   $6.56\pm 0.18$  & $16.01 \pm 0.18$  &  $13.26 \pm 0.25$  \\
 \hline
 $L^1_{norm}$ Res18 = 0 & $6.97\pm 0.15$  & $16.48 \pm 0.12$  &  $14.70 \pm 0.19$  \\
 \hline
 $L^1_{norm}$ Start &     $15.09\pm 0.24$ & $21.08 \pm 0.16$  &  $19.37 \pm 0.15$  \\
 \hline
 Res 18 Voltage [V] &   3.59 & 5.50 &  7.07 \\
 \hline
\end{tabular}
\caption{\textbf{Summary table of the output $L^1_{norm}$ when compensating for the broken resistance}. In the table we report a comparison of the $L^1_{norm}$ when compensating for the broken 18th resistance, training towards the unreachable target obtained with 18 resistances, while using only 17 of the resistances. We report the values for 3 separate trainings, showing in each case the $L^1_{norm}$ between the target matrix and, in order, the trained matrix, the matrix obtained with the target parameters with the 18th resistance turned off, and the matrix of the starting point of the training. We also report the original value of the voltage of the 18th resistance in the target parameters. We can see how, in all cases, the trained chip ($L^1_{norm}$ Trained) is able to perform better than the base configuration ($L^1_{norm}$ Res18 = 0), with a more pronounced effect when the missing heater had a higher voltage. The associated uncertainties are the statistical uncertainties computed with a Monte Carlo procedure. A more in-depth explanation on the method used for computing them is reported in Appendix~\ref{app:ErrorComputation}}
\label{tab:BrokenResComp}
\end{center}
\end{table}

In Tab. \ref{tab:BrokenResComp}, we report the values of the $L^1_{norm}$ between the target sub-matrices and the reconstructed sub-matrices obtained by applying the tensions returned at the end of the trainings, and compare that value of the norm to the one obtained when comparing the target with the matrix reconstructed from using the same parameters for the target, except with the 18th heater turned off (which we will henceforth refer to as the ``base" configuration). We can see that the training method is able to perform better than the base configuration in all cases, with a more pronounced effect when the missing heater had a higher value.  The limiting factor for the quality of the trained result in this case is likely the amount of reconfigurability of the chip we used. While the algorithm is capable of partially compensating for the unused heater, its absence further limits the unitaries reachable by the chip, thus a configuration capable of reproducing the correct unitary is likely to be unreachable by the chip in its current state. Nonetheless, these results show that our training method can successfully find a different set of parameters that partially compensates for the missing heater, and it is likely that a chip with a higher degree of reconfigurability could obtain even better results using our method.

\subsection{{Recalibration for small deviations}}

{As already discussed in Sec. \ref{sec:noise_and_stability}, long term operation of the experimental platform may lead to changes in the experimental parameters, thus potentially reducing the effectiveness of a specific trained setting. Here, we investigate the possibility of partially compensating for this effect by using the training algorithm to perform a recalibration of the circuit programming.}
{More specifically, in some use cases we may be interested in recovering the same final output distribution, more than applying a specific unitary transformation. Thus, for small changes in the state of the experimental setup, we can slightly adjust the chip parameters in order to obtain an output distribution closer to the target one. This can be thought of as a way to compensate for small long term changes of the experimental setup by changing the chip parameters. This operation can be effective for small shifts, since it may not be in general possible to compensate for large modification in the system parameter. Since in this case our aim is to recalibrate the system to obtain a specific output distribution, the proper figure of merit to be used in this case is the loss $\mathcal{L}(\{V_i\})$, defined in Eq. \eqref{eq:loss_func}. }

{To perform this recalibration, we first measured the system at a specific date, obtaining the target output distribution $T^{(17)}_{ref}$. After that, we did not manually recalibrate the system for several days, after which we measured again the target distribution of the chip, obtaining $T^{(17)}_{drift}$. We then started a reduced training loop from the parameters used for $T^{(17)}_{drift}$ and $T^{(17)}_{ref}$, with the goal of finding a new set of parameters that can partially compensate the output distribution, $T^{(17)}_{comp}$, whose loss with $T^{(17)}_{ref}$ is smaller than the one of $T^{(17)}_{drift}$. The reduced training loop started directly from the stage involving both single- and two-photon measurements and consisted of only 68 epochs. }

{Before the training, the loss between $T^{(17)}_{drift}$ and $T^{(17)}_{ref}$ was $\mathcal{L}(\{V_i\}) = 0.2381 \pm 0.0002$. At the end of the training, we measured the output distributions of $T^{(17)}_{comp}$ and computed its loss compared to $T^{(17)}_{ref}$, obtaining $\mathcal{L}(\{V_i\}) = 0.2199 \pm 0.0002$. Here uncertainties are calculated by simulating 1000 new output distribution starting from the measured ones, assuming a Poissonian distribution for each element, and then calculating the standard deviation on the losses calculated on the simulated distributions. Thus, these uncertainties reflect only the statistical dispersion due to the number of counts. As a note, these initial and final losses are not the values that we use during training, but are calculated separately using the same measurement time that we use for reconstructing the matrices, and using all available photon pairs. Thus, they show an increased precision when compared to the values of the loss used during training. }

{These results show a slight but not-negligible improvement, showcasing that our training method is also an effective tool for such situations. Also, in this case we optimized for a loss that takes into account both single- and two-photon distributions, and weights them equally. By using different weights for the different distributions, one could force the training method to perform an unbalanced optimization between the different terms. This choice would depend on the specific use-case for the recalibrated device.}

\section{Discussion}

In this work we trained a 128 mode continuously coupled 3D optical chip to perform target unitaries using machine learning techniques. The proposed method showed good noise resistance, and its effectiveness has been shown in a scenario with a limited number of one- and two-photon measurements. In fact, the full 180 epoch training process takes approximately the same time as the measurement time needed to reconstruct 6-7 unitaries at the sampling rate of our experimental setup.
This is due to many of the optimisations we implemented, such as measuring only a subset of the possible photon pairs each epoch, adopting a two-step training process, and optimising the measuring time per input. This proved that this training method can be successfully applied on an integrated device with a large number of modes without requiring any knowledge on its internal structure, and showed its robustness in a real-world scenario. We showed how it performs in two distinct scenarios. First, we have shown its operation when starting from a point of control parameters that is distant and uncorrelated to the target, a regime that can be used to discover the parameters needed to perform a desired transformation in the general case. Second, we showed how it performs when starting from a previously found set of parameters, compensating for changes in the experimental setup caused by time and external factors, and thus working as a recalibration. 
Furthermore, these results pave the way to devising tools to experimentally characterize the subset of physically implementable unitaries from a waveguide array with a specific lattice geometry and equipped with a limited number and spatial arrangement of the available heaters. Once such large-scale continuously coupled chips become reliably trainable for target operations, their application potential broadens significantly. These platforms can be directly deployed for advanced tasks, including randomness manipulation \cite{DiMicco2026BosonSampling128, Hoch2024BernoulliFactory}, reservoir computing \cite{Yan2024ReservoirComputing}, and solving optimization problems for systems governed by complex physical Hamiltonians \cite{Yang_CCW_Rev}.

This work thus provides a step forward in the ability to program large-scale reprogrammable photonic chips, even in scenarios where a full characterisation of said chips is unfeasible. Being a black-box approach, it is in principle a reliable method to train any photonic architecture. Furthermore, while in this work we used this training method to obtain target unitaries, this methodology can be in general applied to train optical chips to perform any desired operation, since as targets it employs only a part of the output photon distributions of the desired transformation. This means that this training method could also be applied to more complex quantum optical circuits directly on-device, which will become a necessity once these devices surpass the capabilities of classical simulation. As technology advances, we can expect experimental setups with an ever-increasing number of modes, making methods capable of scaling well with the number of modes, ever more important.
However, even though this method has already been shown to scale relatively well with the number of modes \cite{DenisCCSimTrain}, further work is needed to expand this method to scale well with an increased number of trainable parameters. One important optimisation would be to devise a scheme that allows for multiple parameters to be adjusted at the same time, without requiring a significant increase in measurement time per epoch.

\section*{Acknowledgements}
This work is supported by the European Union via project QLASS (“Quantum Glass-based Photonic Integrated Circuits” - Grant Agreement No. 101135876), by PNRIC ‘Q-SUD’ (1\_16868501004\_0001741), Action 1.1.2 ‘Q-SUD’ (1741\_QIIR112\_00051) and by FARE Ricerca in Italia QU-DICE (Grant n. R20TRHTSPA).

\appendix

\section{Correlation between $L^1_{norm}$ and Fidelity}
\label{app:corr_Frob_Fid}

\begin{figure}[h!]
    \centering
    \includegraphics[width=1\linewidth]{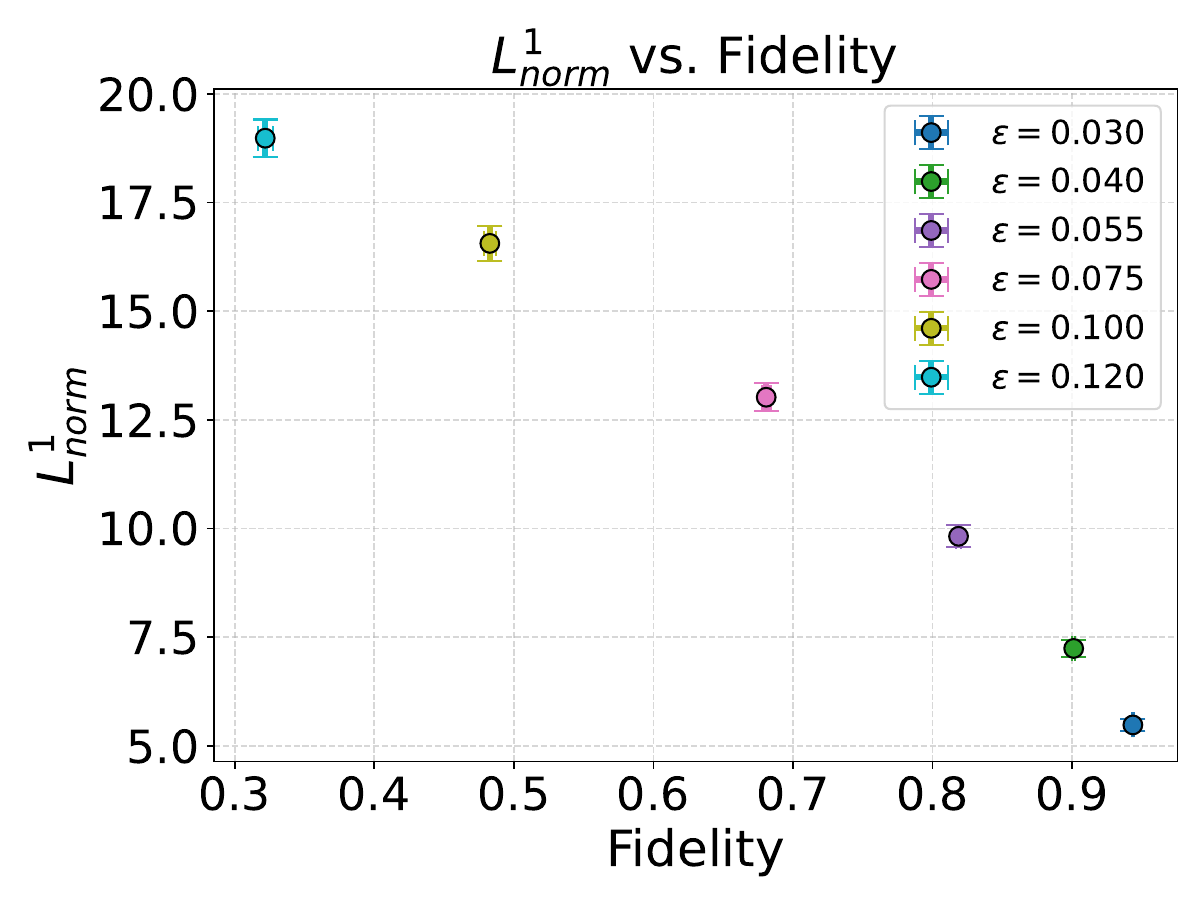}
    \caption{\textbf{Comparison of $L^{(1)}_{norm}$ vs Fidelity}. In the figure we show the comparison between the Fidelities of the complete matrices and the $L^{(1)}_{norm}$ of the respective submatrices for different values of $\epsilon$. Each point is obtained by extracting 1000 different Haar-random unitary matrices, applying a noise $U^{i\epsilon H}$, and comparing the noiseless unitary with the noisy unitary.}
    \label{fig:l1_vs_fidelity}
\end{figure}

In the main text, we make use of the $L^1_{norm}$ of the difference between matrices when comparing the unitaries reconstructed from the trained parameters with the target unitaries. We then use this metric to evaluate the quality of the training output. This section aims to explain the reason of this choice.

Since we do not reconstruct the full unitary, as we only have access to 4 input modes, and 102 output modes, the metric usually employed to evaluate closeness between unitaries, the Fidelity $F = \vert \mathrm{Tr}[(U^{t})^{\dag}  U^{p}] \vert/m$ , cannot be applied. As such, we decided to use a metric that can be applied to these submatrices, but that has a partial correlation with Fidelity.

We have found that, in our use cases, we could find a partial correlation between the Fidelity of the complete unitaries and the $L^1_{norm}$ of the submatrices of said unitaries. We simulated a 128x128 unitary matrix $U_{i}$, and compared it with the same matrix multiplied by a noise unitary $U^{\mathrm{noise}} $ , obtained as
\begin{equation}
U^{\mathrm{noise}} = e^{\imath \epsilon H},
\end{equation}
where $\epsilon$ is a parameter quantifying the noise strength, and $H$ is a random Hermitian matrix. The noisy unitary is found as $U_{i}^{\mathrm{noisy}} = U_{i}U^{\mathrm{noise}}$. We repeated this process for 1000 {different Haar-random} unitaries for each value of $\epsilon$ considered, and found that for specific values of $\epsilon$ the fidelities remained close to a specific value.

After that, we took a $4 \times 102$ submatrix from said unitary matrices, and compared them using the $L^1_{norm}$, and found that we obtained a similar correlation between the value of $\epsilon$ and the resulting $L^1_{norm}$, though with a greater oscillation when compared to the Fidelity. This thus allows us to use the $L^1_{norm}$ as a partial substitute for the fidelity when dealing with these submatrices, and gives us a point of comparison for the values of the $L^1_{norm}$ obtained. In Fig. \ref{fig:l1_vs_fidelity} we can see the comparison between {the} Fidelity and $L^1_{norm}$ {for different values of $\epsilon$}.

As a final note, this comparison assumes that the noise is mostly uniform, so this correlation may not apply in situations where the noise impacts certain modes much more than others, but it can nonetheless serve as a useful point of reference for evaluating the performance of our experimental setup and training method.

\section{Impact of short-term stability on the loss} 
\label{app:LossStability}

\begin{figure}[h]
    \centering
    \includegraphics[width=1\linewidth]{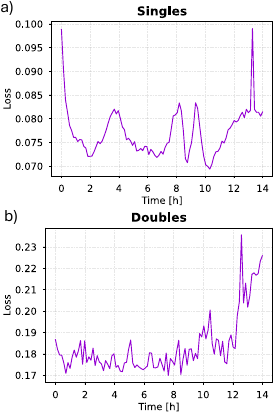}
    \caption{
    \textbf{Stability measurements.} The figures show the evolution of the loss over a $14$ hour measurement period, computed with respect to the target state $T_{ref}^{(17)}$, without modifying any parameter of the experimental setup. Panels \textbf{(a)} and \textbf{(b)} report the losses obtained from single-photon and two-photon measurements, respectively.
    For the single-photon measurements, the output probability distribution was recorded by injecting photons into one input mode at a time. Each acquisition lasted $6~s$. The reported loss corresponds to the average of the four losses associated with the four input modes.
    For the two-photon measurements, the coincidence distribution was recorded by injecting photons into two input modes simultaneously. Each acquisition lasted $60~s$. The reported loss corresponds to the average of the six losses associated with the six possible input pairs.}
    \label{fig:StabilityMeasure}
\end{figure}

In addition to the stability analysis reported in the main text, we further assessed the short-term stability of the system using the loss function defined in Eq.~\eqref{eq:loss_func}, to better understand how these sources of noise impact the training. Fig.~\ref{fig:StabilityMeasure} reports the temporal evolution of the loss for both single-photon and two-photon measurements, obtained without any modification to the experimental setup, and without changing any of its parameters. This allows us to see the fluctuations in the loss caused only by the experimental setup, without the effect of the training algorithm, since the parameters are kept fixed for these measurements. This, in turn, can give us important information to help decide the best value of $\delta_{shift}$ such that the improvement in performance caused by the training step is not obscured by noise, and can also give us an indication of what is the lowest value of the loss that is experimentally reachable. \\
The single-photon loss remains remarkably stable throughout the measurement, apart from a few localised fluctuations that rapidly decay. In contrast, the two-photon loss exhibits significantly larger variations. While several sources of noise contribute to these variations, including drift in the single photon source, the main source of short-term uncertainty for these measurements seem to be statistical fluctuations.
In addition, a significant increase in the loss can be observed during the final stage of the acquisition; in our tests, we found that these sudden spikes in loss did occasionally happen, but that they stabilised after a short while. We speculate that this fluctuation is due to a temporary instability of the single-photon source.

\section{Loss Landscape}
\label{app:LossLandscape}

\begin{figure*}[t]
    \centering
    \includegraphics[width=1\linewidth]{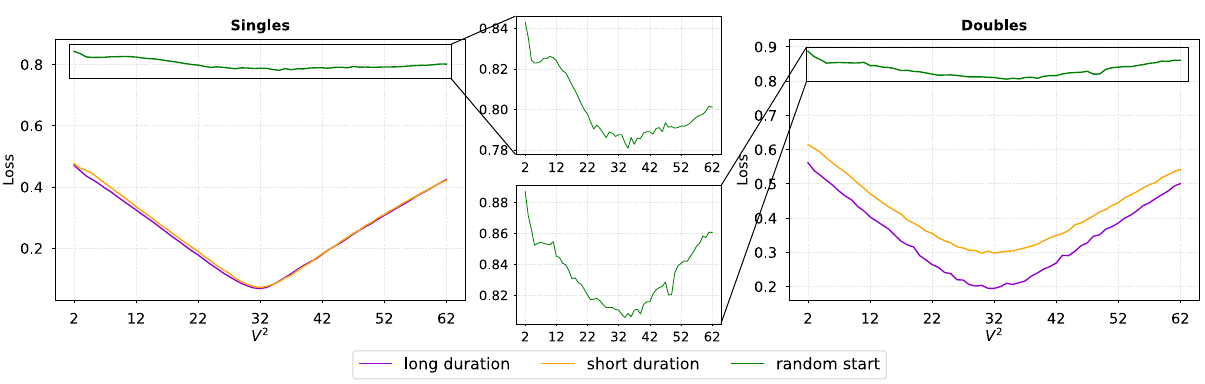}
    \caption{\textbf{Loss Landscape.} In this figure, we show the loss landscape for 1 and 2 photons. The measurements were taken by setting all parameters to a specific value, and then changing only one of the parameters from 2 $V^2$ to 62 $V^2$, in intervals of 1 $V^2$. In purple, the case where the starting parameters were set to the target parameter, and measurements were taken for 6 seconds for the singles, and 60 seconds for the doubles. In orange the case where measurements were taken instead for 1 and 10 seconds respectively. In green the case where the starting parameters were set to significantly different values from those of the target, and measured for 6 and 60 seconds respectively. The insets show a zoom of the green lines. The value of the shifted parameter in the target is around 32 $V^2$.}
    \label{fig:LossLandscape}
\end{figure*}

We report here the loss landscape when shifting a single parameter, which we show in Fig. \ref{fig:LossLandscape}, as it can give some additional insights on how and why the training process works. We show the loss landscape both for the case where only the shifted parameter is different from the target ones, which shows us the correlation between $V_i^2$ and the loss for the single parameter, and the effect of noise when close to the target, and for the case where all parameters are set to random values uncorrelated to the target, which shows how the loss landscape for that parameter is at the beginning of the training. 

To do this loss landscape, we set all parameters to a fixed value, and changed the value of only one of the trainable parameters, going through almost the complete range of values available, at regular intervals. To be more precise, the values of $V_i^2$ we used ranged from 2 $\mathrm{V^2}$ to 62 $\mathrm{V^2}$, with steps of 1 $\mathrm{V^2}$. As we discussed in the main text, we chose to use $V_i^2$ instead of $V_i$ for parameters to obtain a behaviour of the loss that should be close to linear, and with the loss landscape, we can see that this assumption proved to be mostly true. In the figure we can see how the different measurement times affect the loss landscape, seeing how the impact of shorter measurement time on the single-photon measurements is mostly negligible, while it is significantly more pronounced for the two-photon measurements. The loss landscape from the random starting point shows that, while the absolute minimum seems to be close to the correct value, it still presents some local minima, and relatively low oscillations in the loss value. Despite this, our training method is still capable of obtaining satisfactory results in most cases, showing a certain robustness of the method.

\section{{Loss during training}}
\label{app:LossTraining}

\begin{figure}[h]
    \centering
    \includegraphics[width=1\linewidth]{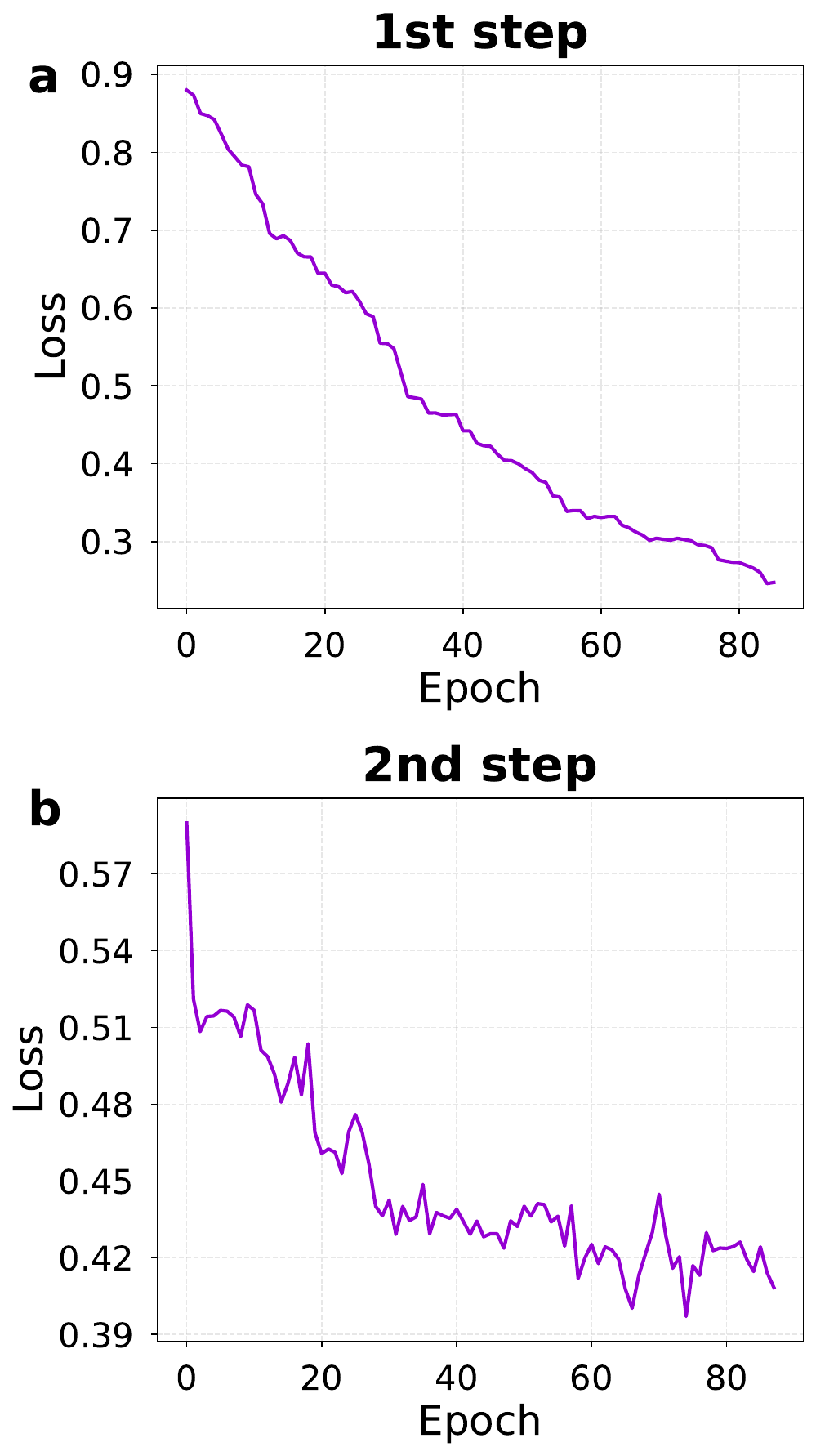}
    \caption{
    {\textbf{Loss during training.} In this figure we can see the behaviour of the loss during the training towards target $T_9$. Panel \textbf{(a)} shows the first step of training, where only single photons were used, while panel \textbf{(b)} shows the second step, where we also used photon pairs for the training.} }
    \label{fig:LossTraining}
\end{figure}

{We report in this section the behaviour of the loss during the training process; in particular,  in Fig. \ref{fig:LossTraining} we show its behaviour during one of the trainings, the one towards the target $T_9$ to be specific, which is representative of its behaviour in most of the trainings.}

{As we can see, the first step of training, which uses only single photons, initially decreases very quickly, with the decrease of the loss slowing down near the end. The second step of training then gives another sharp decrease in the loss in the earlier epochs, after which the decrease in the loss slows down, before stabilising at the later epochs, where the various sources of noise cause fluctuations in the loss that do not allow it to decrease further. As we can see, using the two-step training process does give us the benefits discussed in the main text, since using first the faster single-photon training step allows us to quickly pass through the initial phases of training, where less accuracy is required, while adding the two-photon measurements in the second step allows us to increase the level of precision in the latter epochs.}

{As a point of note, one can see that the starting value for the loss for the second step is higher than the final value of the loss for the first step, even though the second step of training starts from the parameters with which the first step ended. This is because in the second step the loss considers also the two-photon distributions, which were not considered in the first step, thus leading to a higher value of the loss, even though they both represent the same situation.}\\
{\phantom{a-}Overall, the figure shows a decreasing behaviour of the loss that is indicative of an overall good training performance and, since as we saw in the main text the final trained matrix is close to the target one, indicates that our choice of loss was appropriate for the task at hand.}

\section{Computation of statistical uncertainties} \label{app:ErrorComputation}
\begin{figure}[h]
    \centering
    \includegraphics[width=1\linewidth]{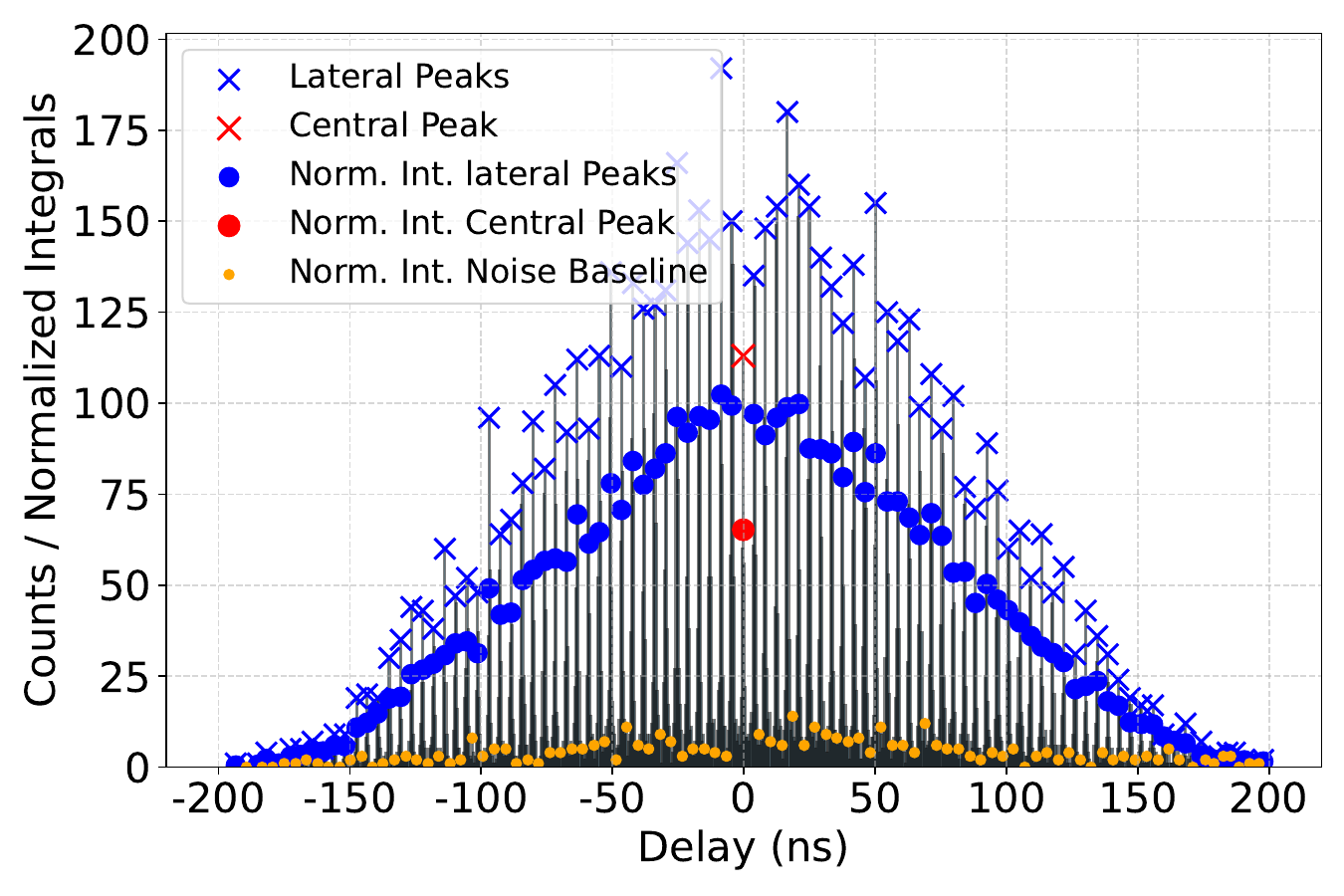}
    \caption{\textbf{Two-photon coincidence histogram.} The figure reports an example of a coincidence histogram used to compute the visibilities $V_{i,j,g,h}$. The linear decrease of the lateral peaks is due to the time-to-space demultiplexing system, which routes photons delivered in trains of 200~ns duration, giving rise to the characteristic triangular shape visible in the figure.}
    \label{fig:corrhist}
\end{figure}

In this section we explain in detail the strategy used to assign uncertainties to the results in Fig.~\ref{fig:norm_target} and Tab.~\ref{tab:BrokenResComp}. The uncertainty on the $L_1$ norms is the statistical uncertainty of the measured quantities, estimated with a Monte Carlo procedure that acts directly on the raw data, before any reconstruction step. The reconstruction algorithm relies on single-photon counts and on the visibility of two-photon coincidences. The single-photon counts are assumed to follow a Poissonian distribution, while sampling the visibilities $V_{i,j,g,h}$ requires a more elaborate treatment. The visibility is defined as the depth of the two-photon coincidence dip: for photons injected in the input modes $i$ and $j$ and detected in the output modes $g$ and $h$, the correlated counts registered at zero delay are subtracted from the uncorrelated background level, and the result is normalized to the latter, so that the quantity is independent of the acquisition time. An example of the acquired histograms is shown in Fig.~\ref{fig:corrhist}. For each fixed pair of input modes, a coincidence histogram is built for every pair of output modes. The correlated counts are those registered at zero delay and are also assumed to be Poissonian. The uncorrelated counts cannot be taken from the heights of the lateral peaks, since the demultiplexing system makes them decrease linearly with their distance from the central peak. Each lateral peak is therefore integrated over a fixed window, and the integrated areas are linearly fitted as a function of peak position, separately on the left and on the right side of the histogram; the uncorrelated level is estimated as the intersection of the two fitted lines, and assumed to be Gaussian-distributed, with standard deviation obtained by propagating the fit covariances through the intersection. The central peak area is likewise sampled from a Poisson distribution. Finally, to suppress the dark-count contribution, the noise baseline between adjacent peaks is fitted with the same procedure, and the corresponding level is subtracted in the visibility formula. Each Monte Carlo realization of the counts and visibilities is then processed through the same reconstruction pipeline used for the nominal data. The uncertainty is then given by the standard deviation of the $L_1$ norm computed over 100 Monte Carlo realizations obtained in this way.

%

\end{document}